\documentclass[a4paper]{jpconf}

\usepackage{graphicx}

\include{babarsym}

\newcommand{\hz}{\ensuremath{h^0}\xspace}
\newcommand{\sinbb}{\ensuremath{\sin2\beta}\xspace}

\newcommand{\sinbbeff}{\ensuremath{\sin2\beta_\mathrm{eff}}\xspace}

\newcommand{\abslambda}{\ensuremath{|\lambda|}\xspace}
\def\calC{{\ensuremath{\cal C}}\xspace}
\def\calS{{\ensuremath{\cal S}}\xspace}
\newcommand{\dm}{\ensuremath{\Delta m}\xspace}
\newcommand{\dt}{\ensuremath{\Delta t}\xspace}
\newcommand{\Btag}{\ensuremath{B_\mathrm{tag}}\xspace}

\newcommand{\A}{\ensuremath{{\cal A}}\xspace}
\newcommand{\Abar}{\ensuremath{\overline{\A}\xspace}}

\newcommand{\Af}{\ensuremath{\A_f}\xspace}
\newcommand{\Abarf}{\ensuremath{\Abar_f}\xspace}

\newcommand{\D}{\ensuremath{D}\xspace}

\begin{document}
\title{Measurement of ${\mathrm \beta}$ in \B decays to charm and 
charmonium in \babar}

\author{Marco Bomben, on behalf of the \babar\ Collaboration}

\address{Universit\`a degli Studi \& I.N.F.N. - Trieste,
 Via Alfonso Valerio 2, 34100 Trieste - Italy}

\ead{Marco.Bomben@ts.infn.it}

\begin{abstract}
In this article we will review recent \babar\ measurements of Unitarity Triangle
 angle $\beta$ in $\B$ meson decays to $charm$ and $charmonium$.

\end{abstract}

\section{Introduction}

Measurements of time-dependent \CP asymmetries in \Bz meson decays, through
the interference between decays with and without \Bz-\Bzb mixing, have
provided stringent tests on the mechanism of \CP violation in the standard
model (SM). The time-dependent \CP asymmetry amplitude equals to \sinbb
 in the
SM if the
\B meson decays to a final states that can be accessed to by both \Bz and \Bzb.
 The angle $\beta =
\mathrm{arg}(V_{cd}V_{cb}^*/V_{td}V_{tb}^*)$ is a phase in the 
Cabibbo-Kobayashi-Maskawa (CKM) quark-mixing matrix~\cite{CKM}. The phase
difference, $2\beta$, between decays with and without \Bz-\Bzb mixing, arises
through the box diagrams in \Bz-\Bzb mixing, which are dominated by the
diagrams with virtual top quark.

In this paper we present a review of recent measurements of $\beta$ at \babar\
experiment at the asymmetric-energy \epem $B$ Factory PEP-II; 
 neutral \B meson decays to {\it charm} and {\it charmonium} studies 
 are reported,
including precision measurements using $\Bz\ra (\ccbar)\Kz$ decays,
 tree decays with penguin pollutions, $\Bz\ra
D^{(*)\pm}D^\mp$, and new decay modes $\Bz\ra D^{(*)0}\hz$ ($\hz = \piz,\,
\eta^{(\prime)},\, \omega$).

\section{Time-Dependent {\boldmath \CP} Asymmetry}

To measure time-dependent \CP asymmetries, we typically fully reconstruct a
neutral \B meson decaying into a \CP eigenstate. 
From the remaining particles in
the event, the vertex of the other \B meson, \Btag, is reconstructed and
its flavor is identified (tagged). The proper decay time difference $\dt=
t_{\CP}- t_\mathrm{tag}$, between the signal \B ($t_{\CP}$) and \Btag
($t_\mathrm{tag}$) is determined from the measured distance between the
two \B decay vertices projected onto the boost axis
and the boost ($\beta\gamma= 0.56$ at PEP-II) of the
center-of-mass (c.m.) system. The \dt distribution 
 is
given by:
\begin{eqnarray}
F_\pm(\dt) & = & \frac{\Gamma e^{-\Gamma|\dt|}}{4} [ 1\mp \Delta w \pm
 \label{eq:fdt}\\
 & & (1-2w) (\eta_f \calS \sin(\dm\dt) - \calC \cos(\dm\dt)) ]\,,  
\end{eqnarray}
where the upper
(lower) sign is for events with \Btag being identified as a \Bz (\Bzb),
 $\eta_f$ is the \CP eigenvalue of the final state,
\dm is the \Bz-\Bzb mixing frequency, $\Gamma$ is the mean decay rate of
the neutral \B meson, the mistag parameter $w$ is the probability of
incorrectly identifying the flavor of \Btag, and $\Delta w$ is the difference
of $w$ for \Bz and \Bzb. In the SM, the parameters $\calS =
\mathrm{Im}\lambda/(1+\abslambda)$ and $\calC =
(1-\abslambda)/(1+\abslambda)$, where $\lambda=
\frac{q}{p}\frac{\Abarf}{\Af}$, and \Af (\Abarf) is the amplitude of \Bz
(\Bzb) decaying to the \CP final state $f$. In the SM, if only one diagram
contributes to the decay process, $\calS= -\sinbb$ and $\calC =0$. A non-zero
value of \calC would indicate direct \CP violation.

Because there can be other diagrams with a different weak phase, the
experimental result of \calS does not necessarily equal to $-\sinbb$. To
separate the measured value from  the standard model \sinbb, we denote the
measured one \sinbbeff.

\section{\boldmath $\Bz\ra(\ccbar)\Kz $}

The  $\Bz\ra(\ccbar)\Kz $ modes are called ``golden modes'' because of
 relatively large (${\cal O}(10^{-4}$--$10^{-5})$)  
branching fractions, low experimental backgrounds and high
reconstruction efficiencies, and small theoretical
 uncertainties~\cite{cck0corr}.

The \babar\ collaboration studied $ J/\psi\KS$ and $J/\psi\KL$, 
$\psi(2S)\KS$, $\chi_{c1}\KS$, $\eta_c\KS$ and $J/\psi
\Kstarz(\KS\piz)$ decay modes to measure \stwob, 
 reconstructing approximately 6900 \CP-odd signal events and
3700 \CP-even signal events from $383\times 10^{6}$ \BB pairs.
The result was $\sinbbeff=
+0.714\pm 0.032\pm 0.018$ and $\calC = +0.049\pm 0.022\pm
0.017$~\cite{Aubert:2007hm}.

Because of the high experimental precision and low theoretical uncertainty,
the result from these modes serves as a benchmark in the SM; any other
measurements of \sinbb that have a significant deviation from it, beyond the
usually small SM corrections, would indicate evidence for new physics.

\section{\boldmath $\Bz\ra D^{(*)\pm}D^{\mp} $}

The decay $\Bz\ra D^{(*)\pm}D^{\mp} $ is dominated by a color-allowed
Cabibbo-suppressed $\b\ra \ccbar \d$ tree diagram. The penguin diagram in the
SM has a different weak phase and is expected to contribute few percent
correction~\cite{xing} to \CP asymmetry. A large deviation in \sinbbeff
from that in $\Bz\ra(\ccbar)\Kz$ would indicate possible new physics
contribution to the loop in the penguin diagram.

The final state $\Dp\Dm$ is a \CP eigenstate so $\calS= -\sinbb$ and $\calC=0$
in the SM when neglecting the penguin contribution.
The final state $D^{*\pm}D^{\mp}$ is not a \CP eigenstate. The decay
amplitudes can have a strong phase difference $\delta$, i.e., 
$\A(\Bz\ra \Dstarp\Dm)/\A(\Bz\ra \Dstarm\Dp) = R e^{i\delta}$. As a result,
the \calS and \calC parameters, ($+$ for $\Dstarp\Dm$ and 
$-$ for $\Dstarm\Dp$) are 
$\calS_\pm = 2 R \sin(2\beta_{\mathrm{eff}}\pm\delta)/(1+R^2)$, and 
$\calC_\pm = \pm (R^2-1)/(R^2+1)$, assuming there is no direct \CP violation. 

\begin{figure}[h]
\centering
\includegraphics[width=0.45\textwidth]{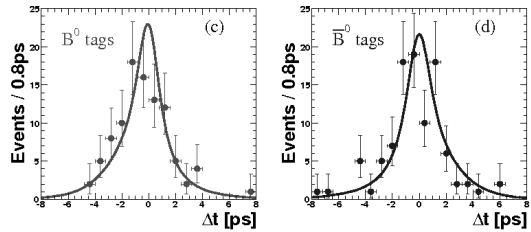}
\caption{\dt distributions of $\Bz\ra\Dp\Dm$.  
    Left plot is  for \Bz-tagged events and
  right one for \Bzb-tagged events. The points represent the data;
  the superimposed is the result of the fit to the data.
\label{fig:DpDm}}
\end{figure}

The \babar\ collaboration measured \CP parameters $\calS$ and $\calC$
 in $\Bz\to\Dp\Dm$ using 384$\times 10^6$ \BB pairs.
The results were: $\calS = -0.54\pm0.34\pm0.06$ and
 $\calC=0.11\pm0.22\pm0.07$~\cite{BabarDpDm}, 
which is consistent with the SM with small penguin contributions. 
Figure~\ref{fig:DpDm} shows the \dt distributions for \Bz-tagged and
\Bzb-tagged events separately. For the result from Belle, the plots show clear
difference between the yields of \Bz-tagged and \Bzb-tagged events. The
consistency of these two results are quite low ($\chi^2/\mathrm{dof}= 12/2$,
or C.L.=0.003, corresponding to $3\sigma$).
 
Belle collaboration recently reports evidence for a large direct \CP violation
in $\Dp\Dm$ channel using a data sample of $535\times 10^{6}$ \BB pairs. They
measure $\calS= -1.13\pm 0.37\pm 0.09$ and $\calC=
-0.91\pm0.23\pm0.06$~\cite{BelleDpDm}. The \CP conservation, $\calS=\calC=0$,
is excluded at $4.1\sigma$ level and $\calC=0$ is excluded at $3.2\sigma$.
The \babar\ and Belle results are marginally compatible: the \chisq
 probability is at 0.3\% Confidende Level (C.L.).

\babar\ also reports an updated measurement of $\Bz\ra \Dstarpm
\Dmp$~\cite{BabarDpDm}. The result is
$\calC_{\Dstarp\Dm} = 0.18\pm 0.15 \pm 0.04$, 
$\calS_{\Dstarp\Dm} = -0.79\pm 0.21 \pm 0.06$,
$\calC_{\Dstarm\Dp} = 0.23\pm 0.15 \pm 0.04$, and
$\calS_{\Dstarm\Dp} = -0.44\pm 0.22 \pm 0.06$. 
Using a slightly different parametrization~\cite{Aubert:2003wr},
 the \babar\ indicated that which indicates that $ \sinbbeff\neq 0$
  it is non-zero at approximately $4\sigma$ level.

\section{\boldmath $\Bz\ra D^{(*)0}\hz$ ($\hz = \piz,\,\eta^{(\prime)},\,\omega$)}

The decay $\Bz\ra D^{(*)0}\hz$ ($\hz = \piz,\,\eta^{(\prime)},\,\omega$) is
dominated by a color-suppressed $\b\ra\c\ubar\d$ tree diagram. The final state
is a \CP eigenstate if the neutral
$D$ meson also decays to a \CP eigenstate, and therefore Eq.~\ref{eq:fdt}
applies. This mode is free of penguin diagrams. The next diagram is also a
color-suppressed tree diagram, $\b\ra\u\cbar\d $, which is doubly Cabibbo
suppressed. The SM correction on \sinbbeff is believed to be a few
percent~\cite{Grossman:1996ke}.
 There could be ``new physics'' contributions thanks to some 
$R$-parity-violating supersymmetry models~\cite{Grossman:1996ke}, entering 
 at tree level.

\babar\ recently reported a measurement of \sinbbeff using $D^{(*)0}\piz$ with
$\Dz\ra\Kp\Km,\,\KS\omega$, and $D^{(*)0}\eta$ with $\Dz\ra\Kp\Km$, and
$\Dz\omega$ with $\Dz\ra\KpKm,\,\KS\omega,\,\KS\piz$. The \Dstarz is
reconstructed from $\Dstarz\ra\Dz\piz$, when applicable. \babar\ uses
$383\times 10^6$ 
\BB pairs and obtains $\sinbbeff = 0.56\pm 0.23\pm 0.05$ and $\calC =
-0.23\pm0.16\pm0.04$~\cite{Aubert:2007mn}. This result is $2.3\sigma$ away
from \CP conservation.

\section{Conclusion}

The measurement of \sinbb is a rich program at the $B$ Factory PEP-II. 
A total of more than 380 million $\FourS\ra\BB$ pairs have been analyzed at
the \babar\ experiment; the result has achieved a precision of 4\% in
 \sinbb measurement,$\sinbb = 0.678\pm0.026$, using
$\Bz\ra(\ccbar)\Kz$ decays.

Belle observes evidence for large direct \CP asymmetry in $\Bz\ra\Dp\Dm$
channel. However, it is not confirmed by \babar, whose result is in 
 agreement with Standard Model expectation $C_f = 0$.
 More data are needed to resolve this discrepancy.
The \babar\ collaboration presented a measurement on  $\Bz\ra\D^{*+}\Dm$ too,
 observing  evidence of \CP violation at 4 standard deviations;
 the results for these analysis are perfectly consistent with the 
 Stadard Model expectations.

The \babar\ collaboration reported the first-time measurement of $S_f$ 
and $C_f$
 in $\Bz\ra D^{(*)0}\hz$ ($\hz = \piz,\, \eta^{(\prime)},\, \omega$) channels,
 when reconstructing $\Dz$ meson in \CP eigenstates; the results are 
 consistent with the Standard Model expectations of $S_f=-\stwob$ and $C_f=0$.

\bigskip 

\end{document}